\documentclass{aa}

\usepackage{graphicx}
\usepackage{txfonts}
\usepackage{natbib}
\usepackage{amssymb}
\usepackage{amsfonts}
\usepackage{amsbsy}
\usepackage{amsmath}
\usepackage{lscape}
\usepackage{algorithm}
\usepackage{xcolor}

\newcommand{\ub}{\boldsymbol{u}}
\newcommand{\xb}{\boldsymbol{x}}
\newcommand{\yb}{\boldsymbol{y}}

\newcommand{\thb}{\boldsymbol{\theta}}
\newcommand{\Rm}{\mathbb{R}}
\newcommand{\Pm}{\mathbb{P}}
 
\begin{document} 

\title{Modeling high-dimensional dependence in astronomical data}
   \author{R. Vio
         \inst{1}
          \and
          T.W. Nagler
        \inst{2}
          \and
          P. Andreani 
         \inst{3}
         }
   \institute{Chip Computers Consulting s.r.l., Viale Don L.~Sturzo 82,
              S.Liberale di Marcon, 30020 Venice, Italy\\
              \email{robertovio@tin.it}
          \and
           Mathematical Institute, Leiden University, Niels Bohrweg 1, 2333 CA Leiden\\
              \email{t.w.nagler@math.leidenuniv.nl}
          \and
                  ESO, Karl Schwarzschild strasse 2, 85748 Garching, Germany\\
                  \email{pandrean@eso.org}
             \email{pandrean@eso.org}    
           }      

   \date{Received....; accepted....}

\abstract{Fixing the relationship of a set of experimental quantities is a fundamental issue in many scientific disciplines. In the 2D case, the classical approach is to compute the linear correlation coefficient $\rho$ from a scatterplot. This method, however, implicitly assumes a linear relationship between the variables. Such an assumption is not always correct. With the use of the partial correlation coefficients, an extension to the 
multidimensional case is possible. However, the problem of the assumed mutual linear relationship of the variables remains. A relatively recent approach that makes it possible to avoid this problem is the modeling of the joint probability density function (PDF) of the data with copulas. These are functions that contain all the information on the relationship between two random variables. Although in principle this approach also can work with multidimensional data, theoretical as well computational difficulties often limit its use to the 2D case. In this paper, we consider an approach based on so-called vine copulas, which overcomes this limitation and at the same time is amenable to a theoretical treatment and feasible from the computational point of view. We applied this method to published data on the near-IR and far-IR luminosities and atomic and molecular masses of the Herschel reference sample, a volume-limited sample in the nearby Universe. We determined the relationship of the luminosities and gas masses and show that the far-IR luminosity can be considered as the key parameter relating the other three quantities. Once removed from the 4D relation, the residual relation among the latter is negligible. This may be interpreted as the correlation between the gas masses and near-IR luminosity being driven by the far-IR luminosity, likely by the star formation activity of the galaxy.}

   \titlerunning{Vine Copulas}
   \authorrunning{Vio, Nagler \& Andreani}
   \maketitle
 
\section{Introduction} \label{sec:intro}

Modeling the relationship of a set of experimental quantities is not straightforward. Often, no theoretical hints are available that would allow us to fix the dependence among the involved variables. Hence, the work has to be entirely based on the analysis of the data.  In the 2D case, an example is represented by the scatterplots and the computation of the corresponding linear correlation coefficients $\rho$.  Its extension to the multidimensional case is possible with the partial correlation coefficients. The main limitation of this approach is the implicit assumption of linear relationships among the variables under study. This is often an unrealistic condition. For this reason, a relatively recent alternative consists of modeling the joint probability distribution function (PDF) of the data. However, this task is not trivial even in the 2D case. Families of bivariate PDFs are available \citep{bal10}, but are not very flexible and are difficult to use. Things worsen for the multidimensional case
\citep{koz00}.  A relatively recent alternative is based on copulas. These are simply multivariate cumulative distribution functions (CDF) with standard uniform margins. They are used to describe
the dependence between random variables, and their main role is to disentangle margins and the dependence structure \citep{nel06, dur16, hof18}. With
copulas it is possible to decompose a joint probability distribution into their margins and a function that couples them. The copula is that coupling function.  

In cosmology, 2D copulas have been used  by \citet{sch10} for the determination of the PDF of the density field of the large-scale structure of the Universe, by \citet{lin15} and \citet{lin16} to predict weak-lensing peak counts, and by 
\citet{sat10, sat11} for the precise estimation of cosmological parameters. Other astronomical applications include the determination of the far-UV and far-IR bivariate luminosity function of galaxies \citep{tak10, tak11},
the determination of the K-band and the submillimeter luminosity function \citep{and14}, and the bivariate luminosity versus the mass functions of the of the local HRS galaxy sample \citep{and18}. 

In principle, the copula approach can work with multidimensional data, but theoretical as well computational difficulties often limit its use to the 2D case. Recently, however, vine copulas have been proposed in the statistical literature as an approach that overcomes this limitation and at the same time is amenable to a theoretical treatment and feasible from the computational point of view.
The strength of vine copulas is that they allow, in addition to the separation of margins and dependence by the copula approach, tail asymmetries and separate multivariate component modeling. This is accommodated by constructing multivariate copulas using only bivariate building blocks, which can be selected independently. These building blocks are glued together to form valid multivariate copulas by appropriate conditioning \citep{joe15, cza19}. This makes  vine copulas a very flexible and reliable tool even in the case of very high-dimensional data.

For this paper, we made use of multidimensional copulas, described in Sects.~\ref{sec:what} and~\ref{sec:prelim}, and in particular of vine copulas, outlined in Sects.~\ref{sec:theory} and~\ref{sec:compv}. We applied them to a data set related to a complete nearby sample of galaxies that has been observed at various wavelengths \citep[][and references therein]{and18} and show its use to highlight the relation to the physical properties of the galaxies in Sect.~\ref{sec:data}.

\section{What are copulas?} \label{sec:what}

A $d$-dimensional copula $C_{1,\ldots,d}(\ub)$, $\ub = (u_1, \dots, u_d) \in [0, 1]^d$ is simply a multivariate CDF with standard uniform univariate margins. Its importance is due to Sklar's theorem: 
for any $d$-dimensional CDF $F(\xb)$, $\xb=(x_1, \ldots, x_d) \in \Rm^d$, with univariate margins $F_1(x_1), \ldots, F_d(x_d)$, a $d$-dimensional copula $C_{1,\ldots,d}(\ub): [0,1]^d \rightarrow [0, 1]$ exists, such that
\begin{equation} \label{eq:C}
F(\xb) = C_{1,\ldots,d}(F_1(x_1), \dots, F_d(x_d))=C_{1,\ldots,d}(u_1, \ldots, u_d),
\end{equation}
where $u_1=F_1(x_1), \ldots, u_d=F_d(x_d)$.
The converse also holds, i.e. given a $d$-dimensional copula $C_{1,\ldots,d}(\ub)$ and univariate CDFs $F_1(x_1), \ldots, F_d(x_d)$, the CDF $F(\xb)$ defined by Eq.~\eqref{eq:C} is a $d$-dimensional CDF with margins $F_1(x_1), \ldots, F_d(x_d)$.
This means that copulas are those functions which combine the univariate margins $F_1(x_1), \ldots, F_d(x_d)$ to form the $d$-dimensional CDF $F(\xb)$. In other words, copulas link multivariate CDFs to their univariate
margins. The importance of copula is more evident if the PDFs $f(\xb)$ are considered. Indeed, it can be shown that
\begin{equation} \label{eq:f}
f(x_1, \ldots, x_d) = c_{1,\ldots,d}(F_1(x_1), \ldots, F_d(x_d)) \cdot f_1(x_1) \cdots f_d(x_d),
\end{equation}
where
\begin{equation} \label{eq:c}
c_{1,\ldots,d}(u_1, \ldots, u_d) = \frac{\partial^d C_{1,\ldots,d}(u_1, \ldots, u_d)}{\partial u_1 \cdots \partial u_d}.
\end{equation}
From Eq.~\eqref{eq:f}, any joint PDF $f(\xb)$ can be factorized into the product of two terms. One is the product of the marginal PDFs $\{ f_i(x_i) \}$  and the other is the copula density $c_{1,\ldots,d}(\ub)$. The first term
provides information on the statistical properties of the individual random variables $\{ x_i \}$ whereas the second term  provides information on their mutual dependence.
Therefore, the importance of $c_{1,\ldots,d}(\ub)$ lies in the fact that it describes the dependence structure among the random variables in separation of the associated marginal PDFs.

If a set of $n$ $d$-dimensional random data $\{ \xb_k \}$, $k=1, \ldots, d$, with $\xb_k=\{x_{p,k}\}$, $p=1,\ldots,n$, is available and the margins $\{ F_k(x_k) \}$ and the corresponding PDFs $\{ f_k(x_k) \}$ are known, the standard procedure to estimate $f(x_1, \ldots, x_d)$ is as follows: first, compute the standard uniform variates  $\ub_k  = F_k(\xb_k)$, then fit their joint distribution by a copula $C_{1,\ldots,d}(\ub| \thb)$, which belongs to a continuous parametric family with characteristic parameters 
$\thb = \{ \theta_1, \ldots, \theta_{n_p} \}$. After that, Eq.~\eqref{eq:f} provides the joint PDF. 
A common method for fitting the copula is based on an estimate of the parameters $\thb$  through a maximum likelihood method, but other techniques are also possible \citep{hof18}. Often, however, the margins are not available. In this case, an alternative is to fit each set $\xb_k$ with a PDF belonging to the Johnson, generalized Lambda, or any other family of parametric PDFs \citep{vio94, kar11} (see also Appendix~\ref{sec:C}),
to compute the uniform random variates $\ub_k  = F_k(\xb_k)$ and then, as before, to fit a copula. When the margins are not estimable with sufficient  accuracy (e.g., because of little available data), a useful nonparametric variant is the computation of the random variates $\ub_k$ by means of the so called pseudo-observations $u_{p,k} = R_{p,k}/(n+1)$ with $R_{p,k}$ the rank of $x_{p,k}$ among $(x_{1,k}, \ldots x_{n,k})$.
Since in general one has no indication of which kind of copula is suited for the data of interest, the typical solution is to fit a set of copulas and to choose that which provides the best result.

In principle, the above procedures can be applied to any $d$-dimensional data set. The point is that most of the parametric copula families available in literature are 2D \citep[e.g., see][]{joe15}, and the few available for a multidimensional analysis are not flexible enough. An alternative approach based on a nonparametric copula estimate has been also proposed \citep[e.g.,][]{nag16, nag17}.

\section{Preliminary considerations} \label{sec:prelim}

Given that most of the available copula families are 2D, it is unclear how a $d$-dimensional PDF $f(x_1,\ldots, x_d)$ can be computed. A possible solution is to express  Eq.~\eqref{eq:f} in terms of 2D copulas. The starting point is that $f(x_1,\ldots, x_d)$ can be factorized into the form
\begin{multline} \label{eq:dec}
f(x_1,\ldots, x_d) = f(x_d) \cdot f(x_{d-1} | x_d) \\ \cdot f(x_{d-2} | x_{d-1}, x_d) \cdots f(x_1 | x_2, \ldots, x_d),
\end{multline}
with $f(x_k | \yb)$ being the conditional PDF of the random variable $x_k$ given the vector of random variables $\yb$.  Now, it can be proved \citep{cza19} that
\begin{equation} \label{eq:fc}
f(x_k | \yb) = c_{x_k y_j | \yb_{-j}}(F(x_k | \yb_{-j}), F(y_j | \yb_{-j}) | \yb_{-j}) \cdot f(x_k | \yb_{-j}),
\end{equation}
where $c_{x_k y_j | \yb_{-j}}(.,.)$ is the conditional copula density, 
\begin{equation} \label{eq:Fc}
F(x_k | \yb) = \frac{\partial C_{x_k,y_j| \yb_{-j}}(F(x_k | \yb_{-j}), F(y_ j| \yb_{-j}) | \yb_{-j})}{\partial F(y_j |  \yb_{-j})},
\end{equation}
$C_{x_k y_j | \yb_{-j}}(.,.)$ is the conditional copula, $y_j$ is one arbitrarily chosen component of $\yb,$ and $\yb_{-j}$ denotes the $y$-vector, excluding this component. The key point is that these conditional PDFs are 
expressed in terms of 2D copula densities. The same holds for the PDF $f(x_1,\ldots, x_d)$. For example, in the 3D case it is
\begin{align} \label{eq:dec3}
f(x_1, x_2, x_3) = & f_1(x_1) \cdot f_2(x_2) \cdot f_3(x_3) \nonumber \\ & \cdot c_{12}(F_1(x_1), F_2(x_2)) \cdot c_{23}(F_2(x_2), F_3(x_3)) \nonumber \\ & \cdot c_{13|2}(F(x_1|x_2), F(x_3|x_2) | x_2).
\end{align}
In fact, the decomposition~\eqref{eq:dec} is not unique since the indices of the variables $\{ x_k \}$ can be permuted. For instance, a decomposition equivalent to ~\eqref{eq:dec3} is
\begin{align} \label{eq:dec4}
f(x_1, x_2, x_3) = & f_2(x_2) \cdot f_1(x_1) \cdot f_3(x_3) \nonumber \\ & \cdot c_{21}(F_2(x_2), F_1(x_1)) \cdot c_{13}(F_1(x_1), F_3(x_3)) \nonumber \\ & \cdot c_{23|1}(F(x_2|x_1), F(x_3|x_1) | x_1).
\end{align}

Although the problem of estimating the PDF $f(x_1,\ldots, x_d)$ has been simplified  by means of Eqs.~\eqref{eq:dec}-\eqref{eq:Fc}, it is still hard to deal with. The conditional copulas $C_{x_k y_j | \yb_{-j}}$ and  corresponding  densities $c_{x_k y_j | \yb_{-j}}$ are difficult to estimate. For this reason, usually the conditional copula densities are simplified into the form 
\begin{equation} \label{eq:simp}
c_{x_k y_j | \yb_{-j}}(F(x_k | \yb_{-j}), F(y_j | \yb_{-j}) | \yb_{-j})  \approx c_{x_k y_j | \yb_{-j}}(F(x | \yb_{-j}), F(y_j | \yb_{-j})).
\end{equation}
Something similarly occurs to the corresponding conditional copulas. This simplification does not only make the problem easier to deal with, but it permits the use of the large set of available continuous parametric families 
of 2D copulas. This makes the method quite flexible. For instance, in the 3D case, the decomposition can be written in the form
\begin{align}
f(x_1, x_2, x_3) = & f_1(x_1) \cdot f_2(x_2) \cdot f_3(x_3) \nonumber \\ & \cdot c_{12}(F_1(x_1), F_2(x_2); \thb_{12}) \cdot c_{23}(F_2(x_2), F_3(x_3); \thb_{23}) \nonumber \\ & \cdot c_{13|2}(F(x_1|x_2), F(x_3|x_2); \thb_{13|2}),
\end{align}
where the 2D copula densities $ c_{12}(.,.; \thb_{12})$, $c_{23}(.,.; \thb_{23})$ and $c_{13|2}(.,.|.; \thb_{13|2})$ can be chosen of different types.

\section{Vine copulas: The theory} \label{sec:theory}
For high-dimensional distributions, there is a huge number of possibilities for decompositions into 2D copulas, named pair-copulas, like Eqs.~\eqref{eq:dec3} and \eqref{eq:dec4}. All these possibilities can be organized according to graphical models called "regular vines". Two special cases, called D-vine and C-vine \citep{aas09}, have been introduced as a simplification. Each model gives a specific way of decomposing a density. 

\begin{figure*}
        \resizebox{\hsize}{!}{\includegraphics{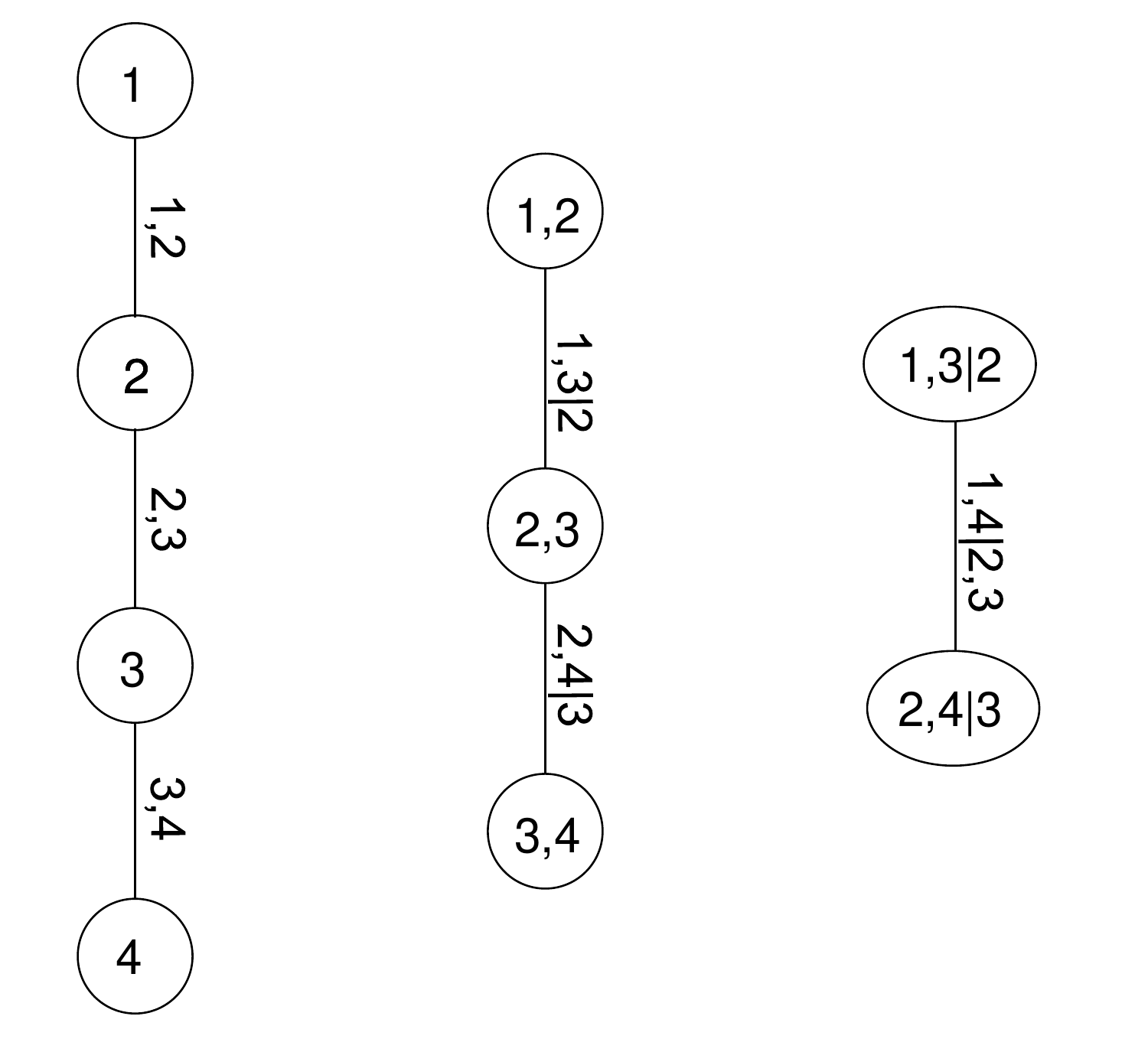}}
        \caption{Example of tree structure of a 4D D-vine copula (see text). }
        \label{fig:fig_d-vine}
    \end{figure*}

Figure~\ref{fig:fig_d-vine} shows the graphical structure of a D-vine for a four-dimensional problem. This structure is formed by three levels or trees.
Each circle or ellipsis constitutes a node, and each pair of nodes is joined by an edge. The label of a node in a given tree is given by the label of the edges of the tree at its immediate left. The label of an edge is given by the indices contained
in the joined nodes with the conditional index given by the common one. For example, in the central tree node $(1,2)$ is connected to node $(2,3)$. The common index is $2$, hence the label of the joining edge
is $(1,3|2)$. Each edge represents a pair-copula density, and the edge label corresponds to the subscript of the pair-copula density. The indices of the CDFs that appear as the argument of a specific pair-copula density are given by the labels of the nodes connected by the corresponding edge. According to this rule, the first tree produces the terms $c_{12}(F_1(x_1), F_2(x_2))$, $c_{23}(F_2(x_2), F_3(x_3))$
and $c_{34}(F_3(x_3), F_4(x_4))$. The second tree produces the terms $c_{13|2}(F(x_1|x_2), F(x_3|x_2))$ and $c_{24|3}(F(x_2|x_3), F(x_4|x_3))$. Finally, the last tree produces the term $c_{14|23}(F(x_1|x_2, x_3), F(x_4|x_2, x_3))$.
The decomposition of $f(x_1, x_2, x_3, x_4)$ is given by the product of these terms:
\begin{align} 
f(&x_1, x_2, x_3, x_4)= f_1(x_1) \cdot  f_2(x_2) \cdot  f_3(x_3) \cdot  f_4(x_4)  \nonumber \\
& \cdot c_{12}(F_1(x_1), F_2(x_2)) \cdot c_{23}(F_2(x_2), F_3(x_3)) \nonumber \\
& \cdot c_{34}(F_3(x_3), F_4(x_4)) \nonumber \\
& \cdot c_{13|2}(F(x_1|x_2), F(x_3|x_2)) \cdot c_{24|3}(F(x_2|x_3), F(x_4|x_3)) \nonumber \\
& \cdot c_{14|23}(F(x_1|x_2, x_3), F(x_4|x_2, x_3)). \label{eq:4d-ex}
\end{align}
For a $d$-dimensional density $f(x_1, \ldots, x_d),$ this procedure provides the decomposition formula
\begin{multline} \label{eq:dv}
f(x_1, \ldots, x_d)  = \prod_{k=1}^d f(x_k) \prod_{j=1}^{d-1} \prod_{i=1}^{d-j} \\ c_{i,i+j|i+1, \ldots,i+ j-1} 
(F(x_i| x_{i+1}, \ldots, x_{i+j-1}), F(x_{i+j} | x_{i+1}, \ldots, x_{i+j-1})),
\end{multline}  
where index $j$ identifies the trees, while index $i$ runs over the edges in each tree.

\begin{figure*}
        \resizebox{\hsize}{!}{\includegraphics{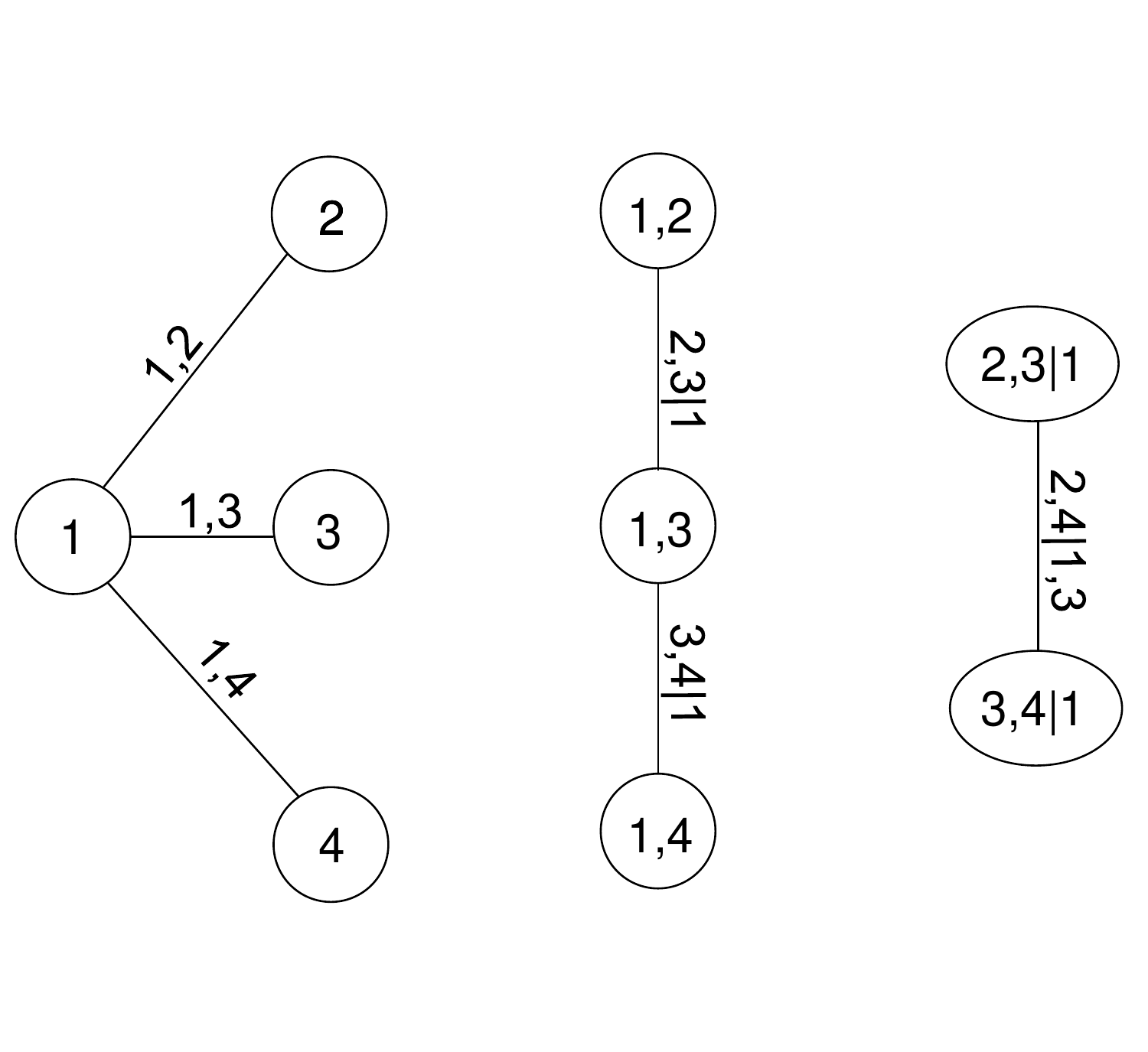}}
        \caption{Example of tree structure of a 4D C-vine copula (see text). }
        \label{fig:fig_c-vine}
    \end{figure*}

Figure~\ref{fig:fig_c-vine} shows the graphical structure of a C-vine again for a 4D problem. While in a D-vine no node in any tree is connected to more than two edges, in a C-vine each tree has a unique node, known as the root node, which is connected 
to all the other nodes. The rules for labeling the nodes and the edges are identical to those of the D-vine. For a C-vine, the decomposition formula is
\begin{align} \label{eq:cv}
f(x_1, \ldots, x_d) = & \prod_{k=1}^d f(x_k) \prod_{j=1}^{d-1} \prod_{i=1}^{d-j} c_{j,i+j|1, \ldots,j-1} \nonumber \\
&(F(x_j| x_{1}, \ldots, x_{j-1}), F(x_{j+i} | x_{1}, \ldots, x_{j-1})).
\end{align}  

Although in principle the decompositions provided by the C-vines and the D-vines should be equivalent, things are actually different  because of the simplification~\eqref{eq:simp}. In general, D-vines are often useful when there is a natural ordering of the variables (e.g., by time), whereas C-vines might be advantageous when a particular variable is known to drive the interactions of the other variables. In such a situation, this variable can be located at the root node of the leftmost tree. In many practical applications, however, no a priori information is available allowing us to decide which kind of vine to use. As a consequence, the decision has to be based on which model better fits the data.

\section{Vine copulas: Computational issues} \label{sec:compv}

The flexibility of vine copulas complicates the parameter estimation and the model selection. One needs to select the appropriate parametric families for each pair-copula, estimate the parameters, and find a good structure for the vine trees. Thankfully, these problems can mostly be solved in separation per pair-copula and per tree level.

We remind the reader that a copula $C_{1,\ldots,d}(\ub)$ is the distribution function of a random variate $\ub  = (u_1, \dots, u_d)$. In what follows, we assume that for all variables $i = 1, \ldots, d$, $n$ uniform variates $\ub_k= \{ u_{p, k} \}$, $p = 1, \ldots, n$, are available. As mentioned in Sect. \ref{sec:what}, these are commonly obtained by transforming the original data $\{ \xb_k \}$  by means of $\ub_k = F_k(\xb_k)$.

\subsection{Model fitting in the 2D case} \label{sec:fitting_bivariate}

We first consider the simpler 2D case. Let us suppose that we have available the variates $\{ (u_{p, 1}, u_{p, 2}) \}$, $p = 1, \dots, n$, from a parametric copula model $c_{12}(u_1, u_2; \thb)$. Then, the parameters $\thb$ can be estimated by  maximum-likelihood: 
\begin{align}
\hat \thb = \arg \max_{\thb} \sum_{p = 1}^n \ln c_{12}(u_{p, 1}, u_{p, 2}; \thb).
\end{align}
In practice, since the true copula is unknown, it is necessary to choose a parametric copula density $c_{12}^{{\mathcal F}_{\kappa}}(.,.)$  from a set of families $\{ {\mathcal F}_{\kappa} \} $, $\kappa  =1, \dots, m$, with $n_{p_k}$ parameters each. This is commonly done by  estimating the parameters  
$ \hat \thb_{\kappa}$ for each copula density and then by choosing the one with  either the lowest Aikaike information criterion (AIC) or the lowest Bayesian information criterion (BIC) (see Appendix~\ref{sec:D}) where \citep{cza19}
\begin{align}
\mathrm{AIC_{\kappa}} &= -2\sum_{p = 1}^n \ln c_{12}^{\mathcal F_{\kappa}}(u_{p, 1}, u_{p, 2}; \hat \thb_{\kappa}) + 2 n_{p_{\kappa}}, \\
\mathrm{BIC_{\kappa}} &= -2\sum_{p = 1}^n \ln c_{12}^{\mathcal F_{\kappa}}(u_{p, 1}, u_{p, 2}; \hat \thb_{\kappa}) + \ln(n) n_{p_{\kappa}}.
\end{align}

\subsection{Iterating through tree levels}

The methods above work for a single pair-copula. It is straightforward to apply them to all pair-copulas in the first tree level, but the same is not true in later tree levels. The reason is that, as is shown by Eq.~\eqref{eq:fc},
the estimate of a $d$-dimensional PDF requires the conditional uniform variates $u_{k|-j}=F(x_k | \yb_{-j})$ and $u_{j|-j}= F(y_j | \yb_{-j}),$
which, however, are not directly available. 

To solve this issue, for the moment we suppose that the tree structure is known and the pair-copulas up to the $(\ell - 1)$th tree level have been fit.
In the $\ell$th tree, pair-copulas have the form $c_{i, j|D}$, where $D$ is a set of $\ell - 1$ variable indices called "conditioning set".
Then there are always edges with indices $(i, r | D \backslash  k)$ and $(j, s | D \backslash  s)$ in the $(\ell - 1)$th tree.\footnote{Symbol $H \backslash  r$ means the set $H$ minus its element $r$.}
With the help of the so called $h$-functions,
\begin{equation} \label{eq:h1}
h_{i|r;D}(u_i | u_r) = \int_0^{u_i} c_{i, r | D \backslash  r}(t, u_r) dt,
\end{equation}
and
\begin{equation} \label{eq:h2}
h_{j|s;D}(u_j | u_s) = \int_0^{u_j} c_{j, s | D \backslash  s}(t, u_s ) dt,
\end{equation}
it can be shown that $c_{i, j|D}(.,.)$ is the joint copula density of the random variables 
\begin{equation} \label{eq:u1}
u_{i|D} = h_{i|r;D}(u_{i | D \backslash  r} | u_{r | D \backslash  r}),
\end{equation}
and
\begin{equation} \label{eq:u2}
u_{j|D} = h_{j|s;D}(u_{j | D \backslash  s} | u_{s | D \backslash  s}).
\end{equation}
Here, the point is that the arguments of the $h$-functions have the same form of the corresponding conditional uniform variables, but the conditioning set has one index fewer. 
Therefore, it is possible to iterate the above equation until $D = \emptyset,$ which corresponds to the first tree, where data are available. 
Because the pair-copulas $c_{i, r | D \setminus r}$ and $c_{j, s | D \setminus s}$ have already been estimated, we can substitute the estimated models in the expressions above. In this way, we can transform data from pair-copulas in one tree into data required for the estimation in the next tree.
For example, we can express $u_{1| 2 3} = F(x_1 | x_2, x_3)$ required in Eq.~\eqref{eq:4d-ex} as
\begin{equation}
    u_{1| 2 3} = h_{1|2; 3}(u_{1|2} | u_{3|2}) = \int_0^{u_{1|2}} c_{13|2}(t, u_{3|2}) dt,
\end{equation}
where 
\begin{align}
    u_{1| 2} &= h_{1|2}(u_{1} | u_{2}) = \int_0^{u_1} c_{12}(t, u_2) dt, \\
    u_{3| 2} &= h_{3|2}(u_{3} | u_{2}) = \int_0^{u_3} c_{23}(u_2, t) dt,
\end{align}
and $u_1 = F_1(x_1), u_2 = F_2(x_2), u_3 = F_3(x_3)$.
The analytical form of the $h$-functions is available for the most common copulas \citep{joe97, sch13}.

\subsection{Finding the tree structure}

The remaining issue is how to select the right tree structure. For a C-vine, we need to specify which variable serves as the root node in every tree. For a D-vine, it is sufficient to specify the order of variables in the first tree. If $d > 4$, there are also structures other than D- and C-vines. 

To select an appropriate structure, the heuristic proposed by \citet{dis13}  can be used. Their idea is to capture the strongest dependencies as early as possible in the tree structure. Here, "strength" is defined as the absolute value of Kendall's $\tau$
(for the definition of this quantity, see Appendix~\ref{sec:A}).
We start in the first tree and compute the (empirical) pair-wise Kendall's $\tau$ for all variable pairs. Then, we choose the tree that maximizes the sum of absolute pair-wise Kendall's $\tau$. We fit pair-copula models for the edges and compute data for the next tree. On these data, we again compute the Kendall's $\tau$ for all possible pairs and select the maximum spanning tree.\footnote{A spanning tree is a subset of a graph, which has all the vertices covered with the minimum possible number of edges.} We continue this way, iterating between structure selection, model fitting, and transforming the data until the whole model is fit.  A summary of the whole procedure is given in Algorithm \ref{alg:stepwise} and implemented in the VineCopula R-package
\citep{nag19}.
\begin{algorithm}[t]
        \caption{Iterative fitting of vine copula models}
        \label{alg:stepwise}
        \vspace{6pt}
        {\bfseries Input:} Observations $\ub_1, \dots, \ub_d$. \\       -----------------------------------------------------------------------------\\
        {\bfseries for}  tree levels $\ell=1, \dots, d-1$:

\begin{enumerate}
        \item   Calculate empirical Kendall's $\tau$ values  $\tau_{i, j|D_e}$ for all possible edges $e = (i,j\mid D_e)$.
        \item Select the spanning tree $E_m$ maximizing $\sum_{e \in E_m} |\tau_{e}|$.
        
        \item {\bfseries for all} $e \in E_m$:
        
        \begin{enumerate}[(i)]
                \item Based on data $\ub_{i_e|D_e},  \ub_{j_e|D_e}$, fit a copula model $c_{i_e, j_e|D_e}$ as in Section \ref{sec:fitting_bivariate}. 
                \item Compute corresponding h-functions  $ h_{i_e| j_e;D_e}$, $ h_{j_e| i_e;D_e}$ using formulas \eqref{eq:h1} and \eqref{eq:h2}.
                \item Set
                \begin{align*}
                \ub_{i_e|D_e \cup j_e} &=  h_{i_e|j_e ; D_e}\bigl( \ub_{i_e|D_e}\big| \ub_{j_e | D_e}\bigr), \\
                \ub_{j_e|D_e \cup i_e} &=  h_{j_e|i_e ; D_e}\bigl( \ub_{j_e|D_e}\big| \ub_{i_e|D_e}\bigr).
                \end{align*}
        \end{enumerate}
\end{enumerate}

        \hspace*{2em} {\bfseries end for}\\
        {\bfseries end for}
\end{algorithm}

\section{Application to an experimental set of data} \label{sec:data}

\subsection{Data set}

We made use of the data published in \citet{and18} and complemented the molecular mass values with additional CO(1-0) line data taken at the NRO 45m antenna at Nobeyama \citep{and20a,and20b}. The data set consists of the {\rm K-band} luminosity, ${\rm L_K}$, the infrared luminosity,  ${\rm L_{FIR}}$, the atomic hydrogen mass, ${\rm M_{HI}}$, and the molecular mass, ${\rm M_{H_2}}$, derived from the CO(1-0) line luminosity toward the volume-limited local galaxy sample, the Herschel reference survey (HRS) \citep{bos10}. The data set is extensively described in \citet{and18} and references therein. Being volume limited, the sample contains all the galaxies above a given threshold of {\rm K-band} luminosity, and the analysis would not be largely affected by a flux selection effect. 

These variables were chosen because they are related to the main overall physical properties of the sample and their relation to the star formation activity in the galaxies. We aimed to investigate the relationship of those properties and derive insights into the driving physical mechanism in their interstellar medium.

\subsection{Data analysis and interpretation} \label{sec:data_analysis}

\begin{landscape}
   \begin{figure}
        \resizebox{\hsize}{!}{\includegraphics{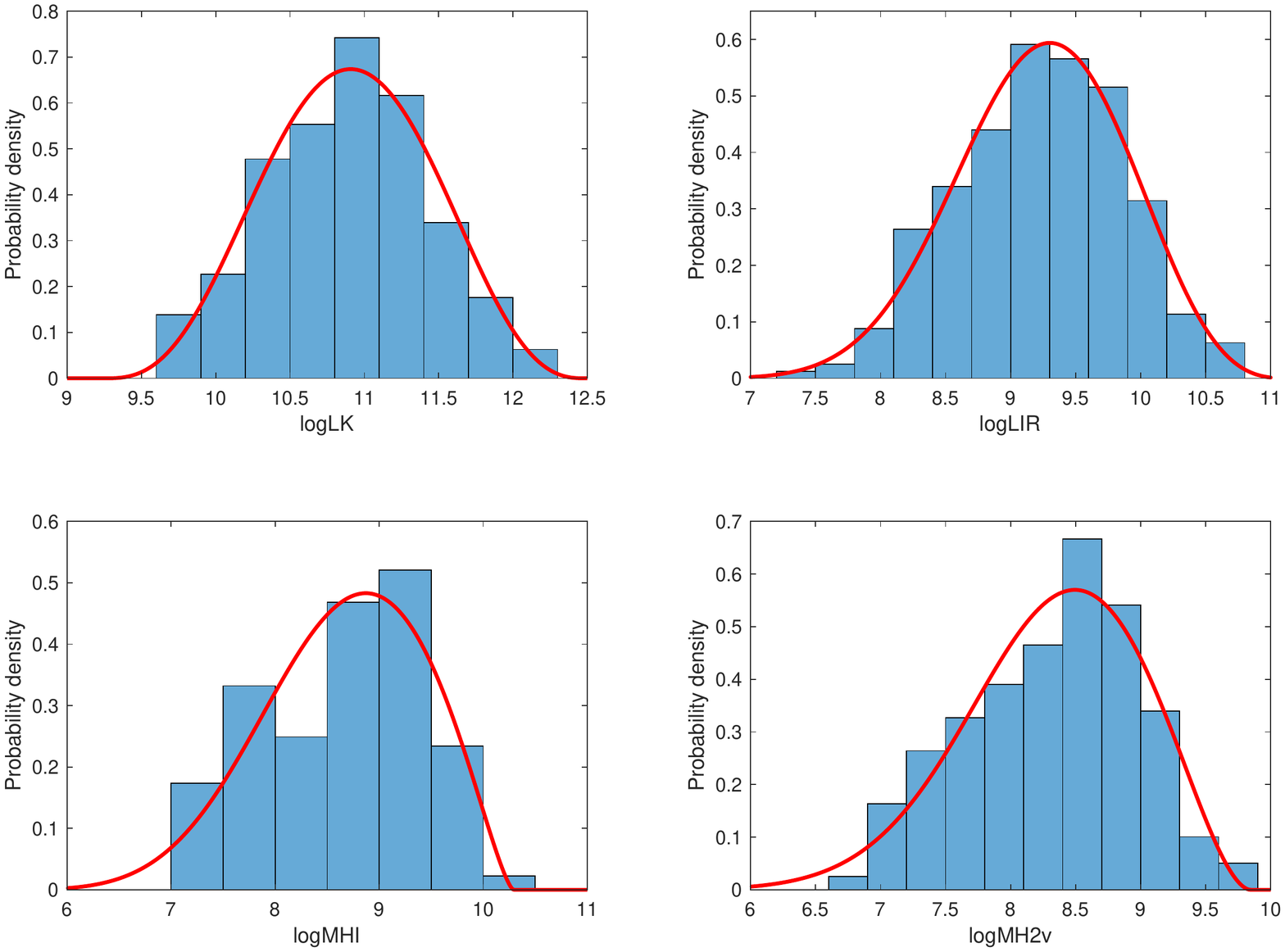}}
        \caption{Histograms of logLK,   logLIR,   logMHI, and   logMH2v data. The red lines provide the PDF obtained by the fit of these data with the generalized lambda distribution family.}
        \label{fig:fig_hists}
    \end{figure}
\end{landscape}

\begin{figure*}
        \resizebox{\hsize}{!}{\includegraphics{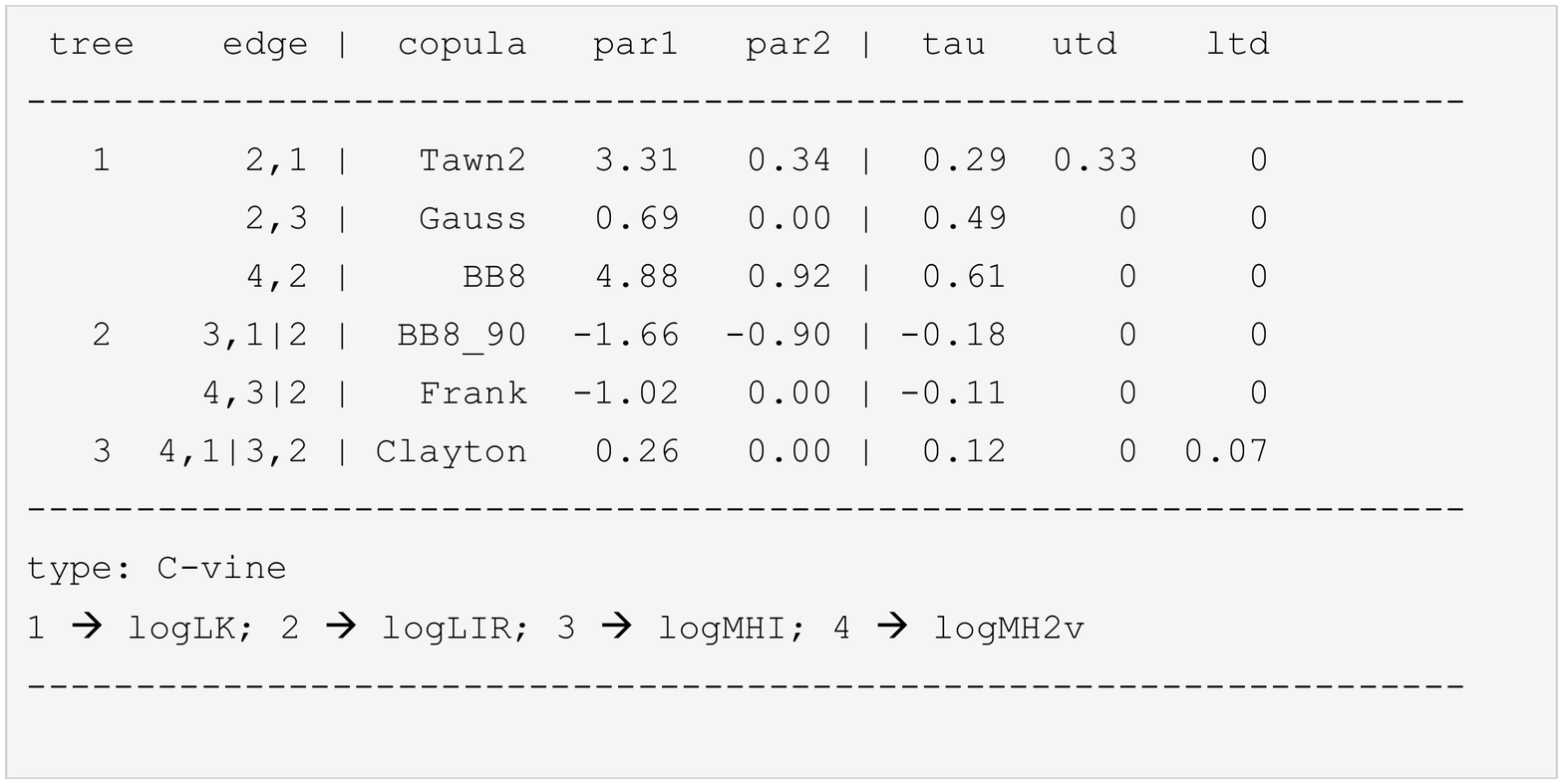}}
        \caption{Results concerning the application of the procedure described in Sects.~\ref{sec:compv} and~\ref{sec:data_analysis} to the data corresponding to Fig.~\ref{fig:fig_hists} with the random variates $\ub$ computed by means of the estimated generalized Lambda PDFs (see text). Here, a copula is associated  to each edge and Kendall's 
$\tau$ for Tree 1, a partial Kendall's $\tau$ for Trees $2$ and $3$ (column "tau") and upper (column "utd"), respectively lower (column "ltd") tail-dependence coefficients. These  are theoretical quantities corresponding to the selected copulas of which the
estimated parameters are given in the columns "par1" and "par2". A description of these copulas can be found in \citet{cza19}. BB8\_90 refers to copula BB8 rotated $90^{\circ}$.}
        \label{fig:fig_results}
    \end{figure*}

\begin{figure*}
        \resizebox{\hsize}{!}{\includegraphics{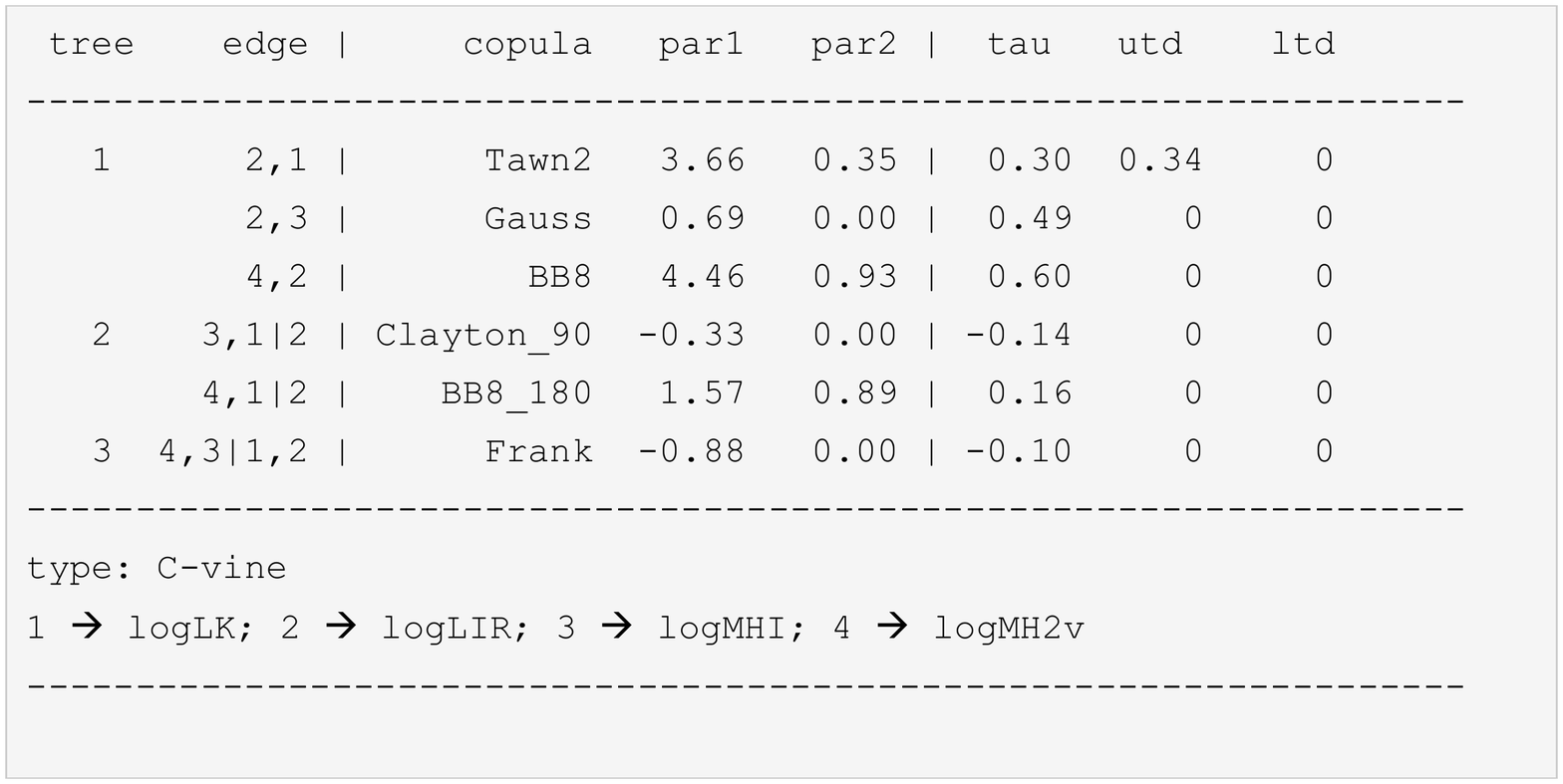}}
        \caption{ As in Fig.~\ref{fig:fig_results}  but with random variates $\ub$ computed by means of the pseudo-observations (see text).  Clayton\_90 and BB8\_180 mean copula Clayton rotated $90^{\circ}$ and copula BB8 rotated $180^{\circ}$, respectively.}
        \label{fig:fig_results_test}
    \end{figure*}

As the first step of the analysis, the PDF of each of the quantities logLK=${\rm log_{10} L_K}$,   logLIR=${\rm log_{10} L_{FIR}}$,   logMHI=${\rm log_{10} M_{HI}}$
and  logMH2v=${\rm log_{10} M_{H_2}}$ were modeled by means of the generalized lambda distribution (GLD) family \citep[][ and references therein]{kar11}. The members of this family are four-parameter PDFs,  which are known for their high flexibility and the large range of shapes that they can reproduce. The starship method has been adopted to fix the parameters. The reason is that this method finds the parameters
that transform the data closest to the uniform distribution, which is an attractive characteristic when working with copulas. The results of the fit are shown in Fig.~\ref{fig:fig_hists}.  After this step, the procedure presented in the previous section was applied
with the random variates $\ub,$ computed by means of the estimated margins.
\begin{table*}
\caption{Sample Kendall's $\tau$.}          
\label{tab:tau}      
%\centering
\hskip -0.7cm          
\begin{tabular}{l c c c c}     % 6 columns 
\hline\hline       
\\
& logLK & logLIR & logMHI & logMH2v \\
\hline    
\\
logLK &   1.00 & 0.25&  0.03 &  0.25\\
logLIR &   0.25 & 1.00 &  0.49 &  0.61\\
logMHI &   0.03 & 0.49 &   1.00 &  0.33\\
logMH2v &   0.25 & 0.61 &  0.33 & 1.00 \\
\\
\hline           
\end{tabular}
\end{table*}
The results are shown in Tab.~\ref{tab:tau} and Fig.~\ref{fig:fig_results}.
The original Kendall's $\tau$ coefficients in Tab.~\ref{tab:tau} are related to the strengths of the relation between the quantities without being dependent on the derived margins. This shows that the strongest correlations occur between the far-IR luminosity ${\rm L_{FIR}}$
and the gas masses, first molecular ${\rm M_{H_2}}$ and then atomic ${\rm M_{HI}}$, while ${\rm L_{FIR}}$ is weakly correlated with the near-IR {\rm K-band} luminosity, ${\rm L_{K}}$.
On the other side, Fig.~\ref{fig:fig_results} indicates the type of vine structure selected, specifically a C-vine, 
and provides the list of pair-copulas singled out for each edge. For each pair-copula, the values of the corresponding coefficients and of the lower and upper tail dependence coefficients are also shown (for the meaning of last two quantities see Appendix~\ref{sec:B}). The Kendall's $\tau$ (for Tree $1$) and partial Kendall's $\tau$ (for Trees $2$ and $3$) associated to each edge are also shown.
This last quantity measures the dependence between two variables after the effect of other variables (the common indices of two nodes) has been removed.\footnote{The “partial Kendall's $\tau$” is computed by means of the Kendall's $\tau$ between the variates $u_{i|D}$and $u_{j|D}$
in Eqs.~\eqref{eq:u1} and \eqref{eq:u2}. It provides a measure of the relationship between $u_i$ and $u_j$ when the influence of the variates corresponding to the set $D$ is removed.} In order to check the reliability of the obtained results, the procedure was repeated with the random variates $\ub$ given by the pseudo-observations. As the comparison of Fig.~\ref{fig:fig_results_test} with Fig.~\ref{fig:fig_results} shows, the differences are not substantial.
The fact that for the highest trees the copulas selected by the two methods are different is not significant. Indeed, one has to take present that when the Kendall's $\tau$ between the random variates coming from two PDFs or two conditional PDFs is close to zero (i.e., they are almost uncorrelated), there are various kinds of copulas that can provide similar reconstructions of the corresponding bivariate data distribution. In other words, in the reconstruction of a multivariate distribution, the specific types of copula are only meaningful for values of the Kendall's $\tau$ significantly different from zero.

\begin{figure*}
        \resizebox{\hsize}{!}{\includegraphics{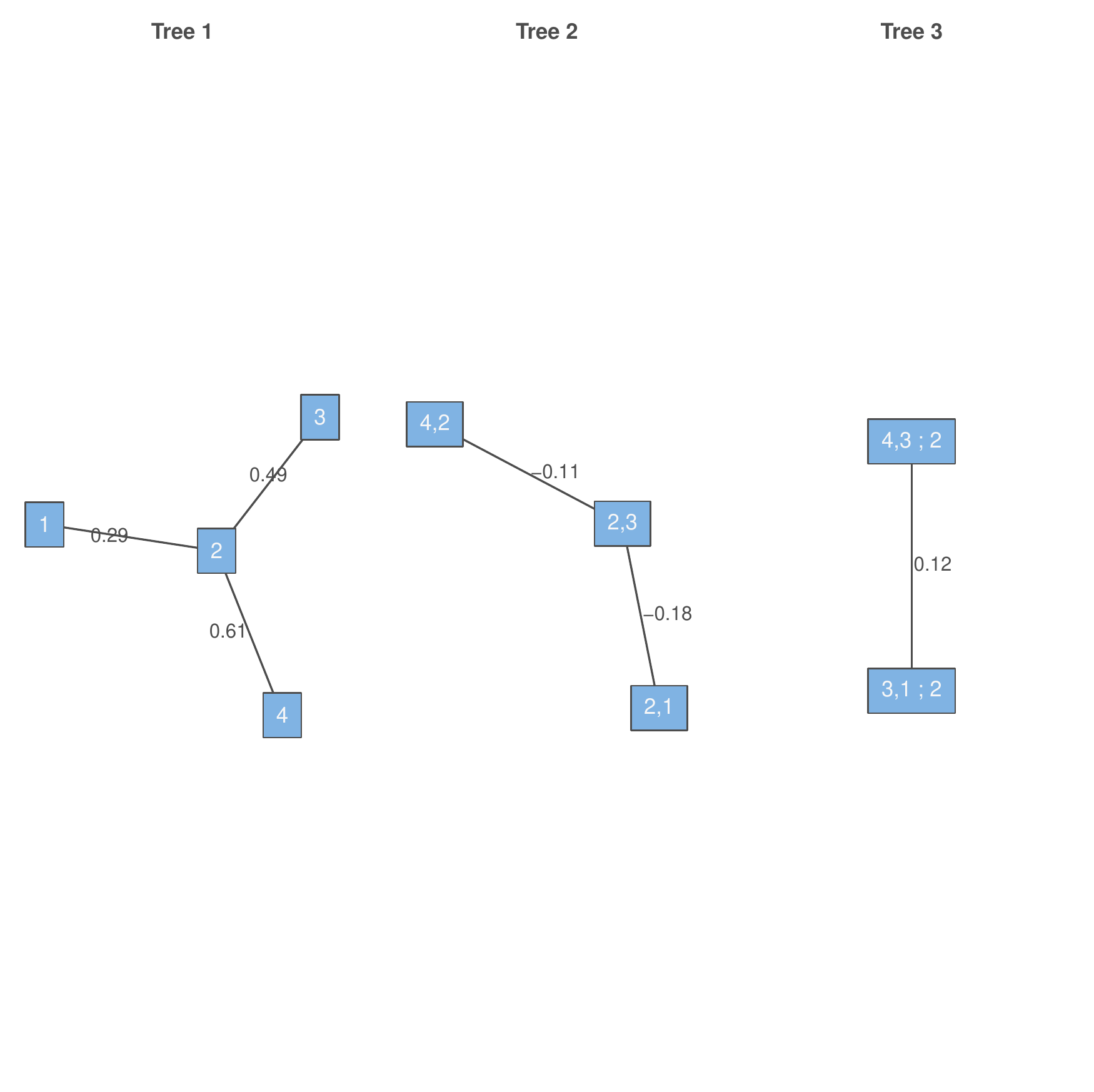}}
        \caption{C-vine structure selected by the procedure described in Sects.~\ref{sec:compv} and~\ref{sec:data_analysis} to the data corresponding to Fig.~\ref{fig:fig_hists}.  Each edge in Tree $1$ is associated to a Kendall's $\tau,$ whereas for Trees $2$ and $3$ they are associated to a partial Kendall's $\tau$. Here, $1 \rightarrow$  logLK, $2 \rightarrow $  logLIR, $3 \rightarrow$  logMHI, $4 \rightarrow$  logMH2v.}
        \label{fig:fig_tree}
    \end{figure*}

These results can be more clearly interpreted by looking at Fig.~\ref{fig:fig_tree}. As explained in Sect.~\ref{sec:compv},
the structure selection algorithm tries to capture the strongest dependencies first. Figure~\ref{fig:fig_tree} shows a plot of the tree structure labeled with the Kendall's $\tau$ (for Tree $1$) and partial Kendall's $\tau$ (for Trees $2$ and $3$). Here, the logLIR quantity as been selected as root node. This means that it is strongly correlated to all other variables and that it drives part of the dependence between the other variables. In the second tree, the effect of logLIR on the dependence between the others has been removed. There is only some weak negative dependence left.

All this can be interpreted with the fact that although from Tab.~\ref{tab:tau} the quantities ${\rm M_{HI}}$ and ${\rm M_{H_2}}$ appear positively dependent, such dependence appears to be driven entirely by their dependence on the quantity ${\rm L_{FIR}}$. 
Once the dependence of ${\rm L_{FIR}}$ is removed from the relation with the other quantities the residual relations 
 ${\rm M_{HI}}$ with ${\rm M_{H_2}}$ and  ${\rm L_{K}}$ with ${\rm M_{HI}}$ are negatively dependent (albeit this dependence is quite weak).
This means that the dependence shown in Tab.~\ref{tab:tau} is driven entirely by their dependence on ${\rm L_{FIR}}$. 

Since  ${\rm L_{FIR}}$ is dominated by the thermal dust emission heated by FUV photons by massive stars and residing in molecular clouds, which are cocoons of star formation processes, this result confirms that the physical properties of the galaxies are driven by their star formation.

\begin{landscape}
   \begin{figure}
        \resizebox{\hsize}{!}{\includegraphics{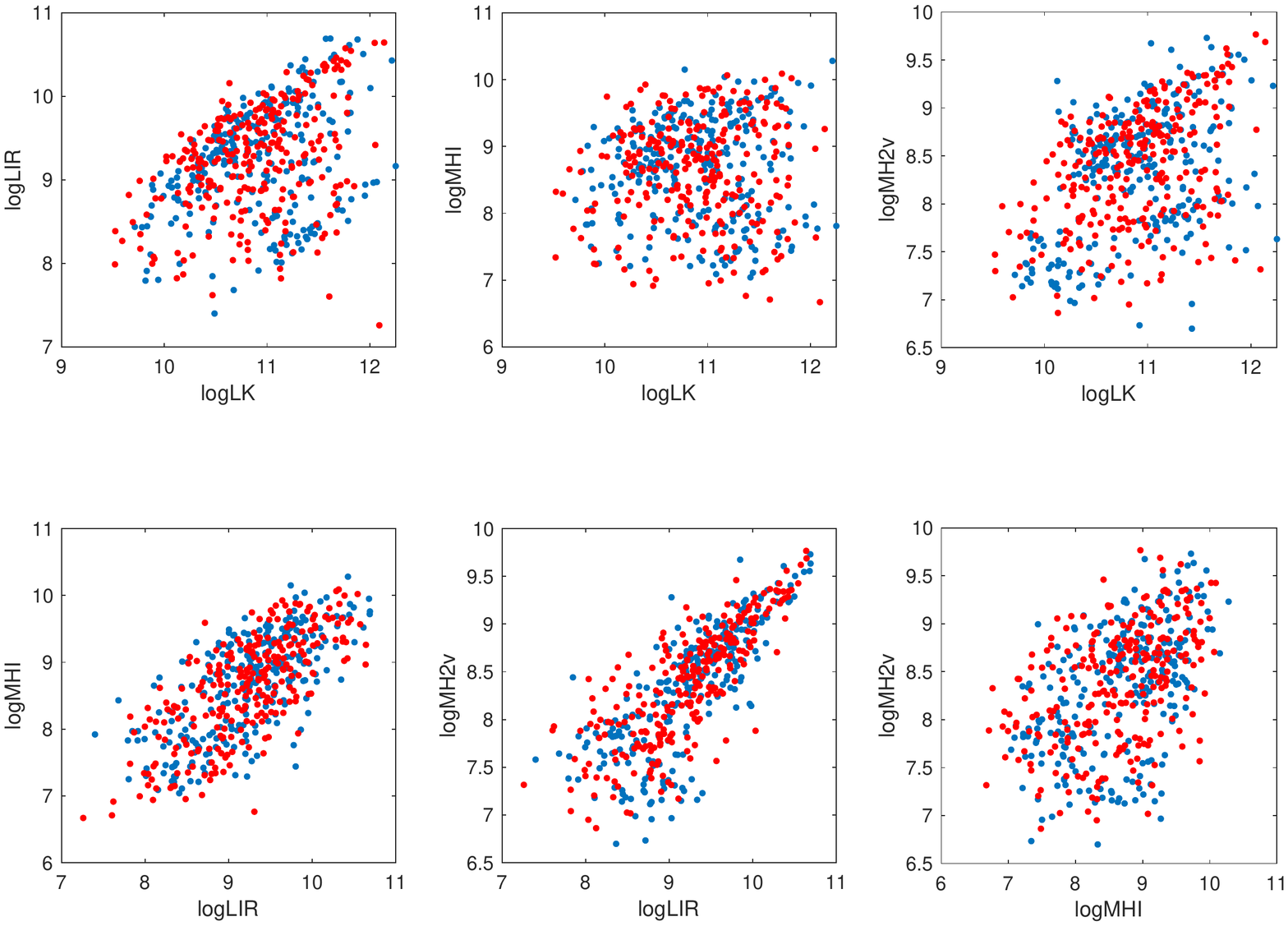}}
        \caption{Original logLK,   logLIR,   logMHI, and   logMH2v data (blue circles) versus the corresponding simulated data obtained from the estimated 4D joint PDF (red circles).}
        \label{fig:fig_ori_sim}
    \end{figure}
\end{landscape}

\begin{figure*}
        \resizebox{\hsize}{!}{\includegraphics{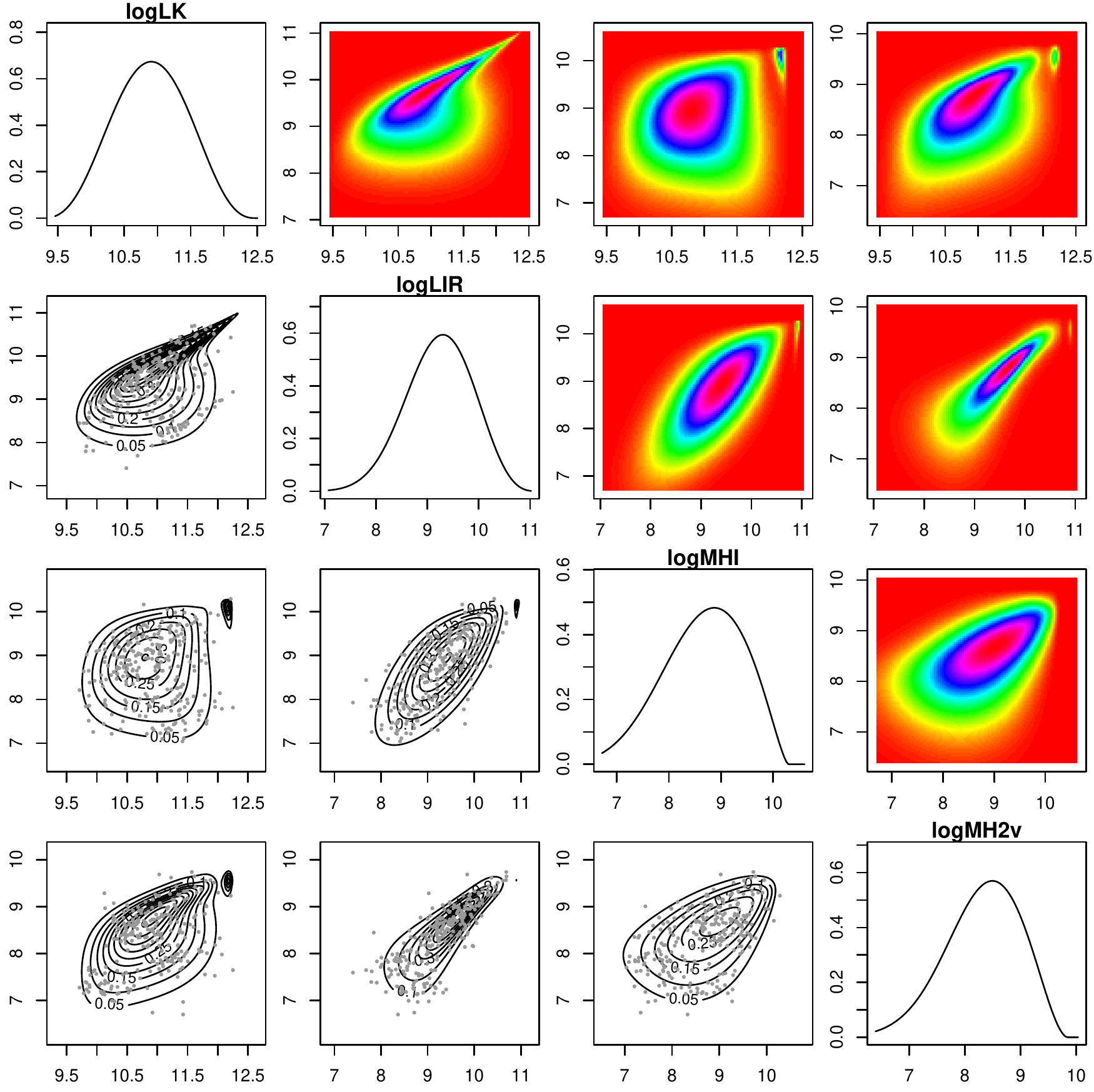}}
        \caption{Two-dimensional slices of the estimated 4D joint PDF for the data corresponding to Fig.~\ref{fig:fig_hists}. Along the diagonal are the fit PDFs of Figure~\ref{fig:fig_hists}. The right panels show the slices in which colors correspond to the intensity of the relation, while the left panels report, on the same slices, the data points and the iso-contours.} 
        \label{fig:fig_slices}
    \end{figure*}

For completeness, in 
Fig.~\ref{fig:fig_ori_sim} we show the original data versus the data simulated from the the estimated 4D joint PDF of which the 2D slices are shown in Fig.~\ref{fig:fig_slices}. Figure~\ref{fig:fig_ori_sim}  shows a good agreement between original and simulated data, while Fig.~\ref{fig:fig_slices} demonstrates the one-to-one relation between the couple of variables.

\section{Conclusions}  \label{sec:conclusions}

In this work, a flexible and effective approach to modeling the relationship of a set of experimental multidimensional quantities is presented. This approach consists
of modeling the joint PDF of the data  by means of a special type of copula called a vine copula. Classical copulas are functions that contain all of the information on the relationship between two random quantities. 
Their major
limitation  is that they are unable to model multidimensional data. Vine copulas overcome this limitation by expressing the joint PDFs as the product of a set of 2D copula densities and the 1D PDFs corresponding to each quantity.  In particular, two types of vine copulas have been considered: the C-vine and the D-vine. This approach makes the estimation of the joint PDFs amenable to a theoretical treatment and feasible from the computational point of view. 

We applied this method to published data on the near-IR and far-IR luminosities and atomic and molecular masses of the HRS. 
We find that the far-IR luminosity, ${\rm L_{FIR}}$, is the key player in driving the galaxy properties in this sample. Despite its original selection in the {\rm K-band,} the HRS sample shows that it is ${\rm L_{FIR}}$ that plays a fundamental role. Removing its dependence from the other variables, the {\rm K-band} luminosity, and the atomic and molecular masses, makes it clear that the established relation among these quantities does not show up any more. 

The ${\rm L_{FIR}}$ in this sample is dominated by the thermal dust emission heated by FUV photons produced by massive stars in molecular clouds. Our analysis therefore highlightsthat the star formation activity of these galaxies is the key parameter driving the galaxy evolution.
\begin{acknowledgements}

\end{acknowledgements}

% here the figures

\clearpage

\clearpage

\clearpage

\clearpage

\clearpage

\clearpage

\clearpage

\clearpage

\clearpage
\appendix

\section{The Johnson's distribution and the generalized Lambda distribution families} \label{sec:C}

The Johnson's distributions and the generalized Lambda distributions (GLD) are both four-parameter families that are used for fitting distributions to a wide variety of data sets. In particular, the Johnson's system is based on three different PDFs, $f_U(x)$, $f_B(x),$ and $f_L(x)$ according to the fact that the random variable $x$ is unbounded, bounded both above and below, and bounded only below:
\begin{multline}
f_U(x)= \frac{\eta}{\sqrt{2 \pi [(x-\epsilon)^2+\lambda^2]}} \\
\times  \exp{\left\{ -\frac{1}{2}  \left[ \gamma + \eta  \ln{\left( \frac{(x-\epsilon)}{\lambda} + \sqrt{\left(\frac{x-\epsilon}{\lambda}\right)^2+1}\right)} \right]^2 \right\}},
\end{multline}
for $-\infty < x < \infty$,
\begin{multline}
f_B(x)= \frac{\eta \lambda}{\sqrt{2 \pi} (x-\epsilon) (\lambda - x + \epsilon)}  \\
\times \exp{\left\{ -\frac{1}{2} \left[ \gamma+ \eta \ln{\left( \frac{x-\epsilon}{\lambda-x+\epsilon}\right)} \right]^2 \right\}},
\end{multline}
for $\epsilon \leq x \leq \lambda+\epsilon$, and
\begin{equation}
f_L(x)= \frac{\eta}{\sqrt{2 \pi} (x-\epsilon)}
\times \exp{\left\{ -\frac{1}{2} \left[ \gamma+ \eta \ln{\left( \frac{x-\epsilon}{\lambda}\right)} \right]^2 \right\}},
\end{equation}
for $x \geq \epsilon$. In literature, various methods are available for the selection of the appropriate type of PDF as well for the estimate of the parameter \citep[e.g.,][]{vio94, kar11}.  

The PDFs corresponding to the GLD family are given by
\begin{equation}
f_{\lambda}(x) = \frac{\lambda_2}{\lambda_3 y^{\lambda_3 - 1} + \lambda_4 (1-y)^{\lambda_4-1}},
\end{equation}
where $x=Q(y; \lambda_1, \lambda_2, \lambda_3, \lambda_4)$ with
\begin{equation}
Q(y; \lambda_1, \lambda_2, \lambda_3, \lambda_4)=\lambda_1 + \frac{y^{\lambda_3}-(1-y)^{\lambda_4}}{\lambda_2},
\end{equation}
and $ 0 \leq y \leq 1$.
Concerning this family, various methods are also available for the estimate of the parameters \citep[e.g.,][]{kar11}.

\section{AIC and BIC criteria} \label{sec:D}

The Akaike information criterion (AIC) and the Bayesian information criterion (BIC) are two criteria for model selection from a finite set of models \citep{bur02}. They are based on the maximum value $\hat{L}$ of the likelihood function for the model as well on the number $n_p$ of free parameters it contains. The idea is that, when fitting models, it is possible to increase the likelihood by adding parameters, but doing so may result in overfitting. Both the BIC and AIC attempt to resolve this problem by introducing a penalty term for the number of parameters in the model. In particular,
\begin{equation} \label{eq:AIC}
{\rm AIC} = 2 n_p - 2 \ln{(\hat{L})},
\end{equation}
whereas
\begin{equation} \label{eq:BIC}
{\rm BIC}=n_p \ln{(n)} - 2 \ln{(\hat{L})}
,\end{equation}
with $n$ being the number of data. In practical applications, a set of models is chosen, the corresponding quantity $\hat{L}$ evaluated, Eq.~\eqref{eq:AIC}  or Eq.~\eqref{eq:BIC} used, and finally the model with the lowest AIC or BIC selected. 
The difference between AIC and BIC is how much model complexity (i.e., the number of parameters) is penalized. For $n \ge 8$, the BIC penalty is stronger. Both criteria give a mathematical guarantee to find the "best" model as the sample size increases. The BIC assumes that the true model is among the set of candidates, but the AIC does not. 
These criteria are useful in the context of the vine copulas since the different types of bivariate copulas considered for their construction contain a different number of free parameters.

\section{The Kendall's $\tau$} \label{sec:A}

\begin{figure*}
        \resizebox{\hsize}{!}{\includegraphics{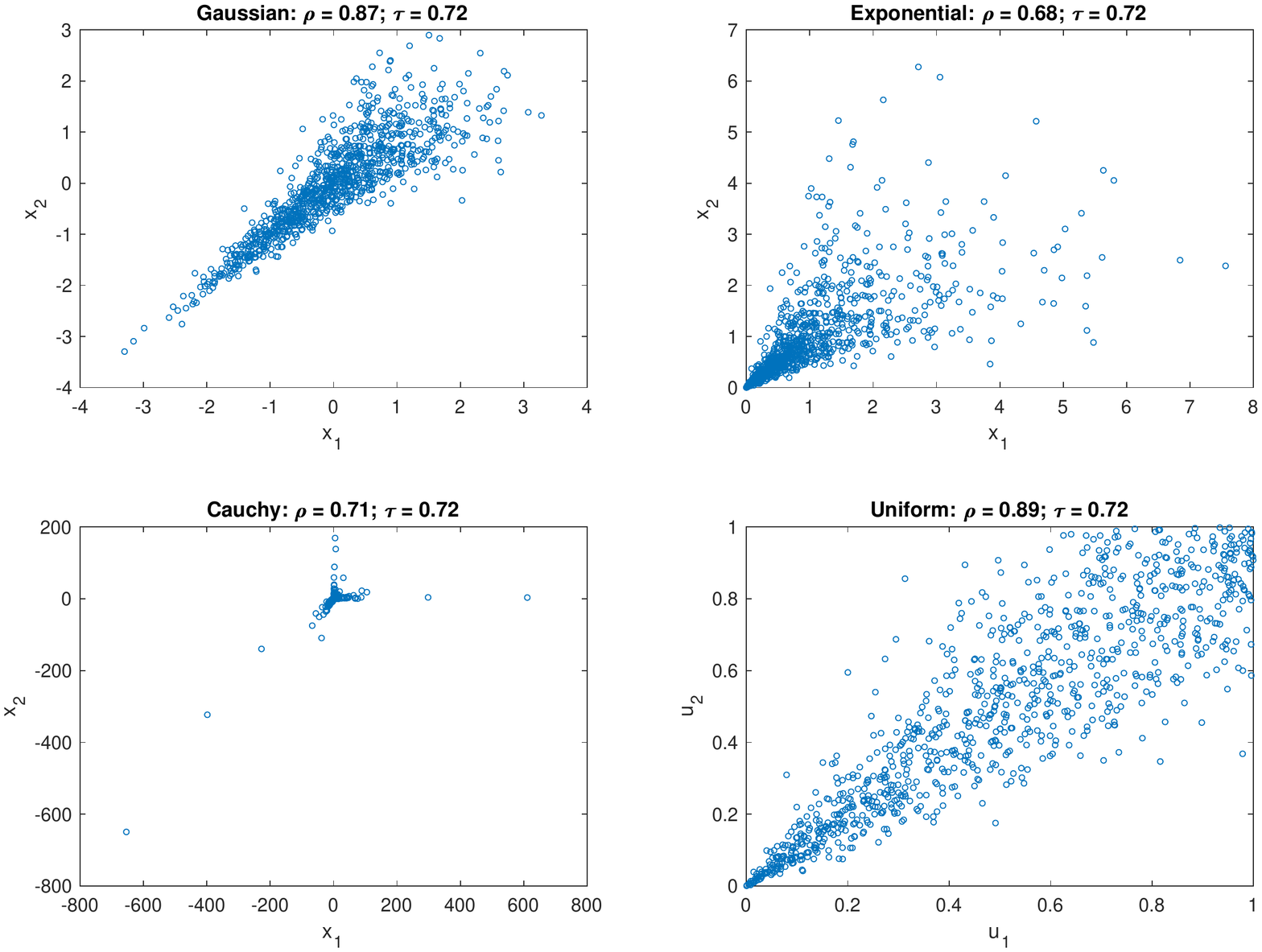}}
        \caption{Numerical realization of $1000$ independent copies of a bivariate random vector $(x_1, x_2)$ from the Gaussian, exponential, and Cauchy PDFs obtained from the set of uniform random pairs $\{ (u_{1,i}, u_{2,i}) \}$, shown in the bottom-right 
panel, by means of the transformations $(x_1, x_2) = (F^{-1}(u_1), F^{-1}(u_2))$ where $F^{-1}(u)$ is the inverse CDF corresponding to the various PDFs.}
        \label{fig:fig_ktau}
    \end{figure*}

When working with copulas the relationship between two random quantities is typically measured by means of the Kendall's $\tau$. The reason can be understood by looking at Fig.~\ref{fig:fig_ktau}, which shows the the realization of $1000$ 
independent copies of a bivariate random vector $(x_1, x_2)$ from the Gaussian, exponential and Cauchy PDFs and of the same number of a bivariate random vector $(u_1, u_2)$ from the uniform PDF. These  realizations appear quite different from one another, as well as the corresponding linear correlation coefficients $\rho$.
Here, the point is that the first three sets of random numbers $\{(x_{1,i}, x_{2,i}) \}$ were obtained from the set of uniform random pairs $\{ (u_{1,i}, u_{2,i}) \}$ by means of the transformations:
\begin{equation}
(x_1, x_2) = (F^{-1}(u_1), F^{-1}(u_2)),
\end{equation}
where $F^{-1}(u)$ is the inverse CDF corresponding to the various PDFs. This is a common method to simulate random numbers from a given PDF. What this figure indicates is that the different appearance of the realizations is not due to the intrinsic relationship between the random quantities, but rather to their margins. Since with copulas one wants to disentangle margins from the dependence structure, the latter should be measured in a way that does not depend on the marginal distributions.
This is what the Kendall's $\tau$ does.

 If $(x'_1, x'_2)$ is and independent copy of $(x_1, x_2)$, $\tau$ is defined as 
\begin{equation}
\tau=\Pm\left[(x_1-x'_1) (x_2-x'_2) > 0 \right] -  \Pm\left[(x_1-x'_1) (x_2-x'_2) < 0 \right],
\end{equation}
that is, it is the probability of concordance minus the probability of discordance of the random pairs $(x_1,x_2)$ and $(x'_1,x'_2)$. The rationale behind this definition is that if there is positive dependence between the variable $x_1$ and $x_2$, then when $x_1$ increases or decreases, a similar behavior has to be expected for $x_2$. It can be demonstrated \citep{hof18} that
\begin{equation}
\tau=4 \int_0^1 \int_0^1 c(u_1, u_2) C(u_1, u_2) d u_1d u_2 -1,  
\end{equation}
meaning that $\tau$ effectively depends only on the underlying copula.

The sample version $\hat{\tau}$ of $\tau$ is given by
\begin{equation}
\hat{\tau} = \frac{2}{n (n-1)} \sum_{i=1}^{n-1} \sum_{j=i+1}^n {\rm sign}[(x_{i1}-x_{j1}) (x_{i2}-x_{j2})],
\end{equation}
where $n$ is the number of observations and ${\rm sign}[x] =1$ if $x > 0$, ${\rm sign}[x] =0$ if $x=0$ and  ${\rm sign}[x] =-1$ if $x < 0$. As expected, $\hat{\tau}$ is the same for all the realizations in Fig.~\ref{fig:fig_ktau}.

\section{Tail-dependence coefficients} \label{sec:B}

There are situations where in the 2D scatterplot of a set of data, the points appear concentrated in one or both the tails of their joint distribution. 
For instance, this is the case for the scatterplots in Fig.~\ref{fig:fig_ktau}, where a concentration of points in the lower-left tail of the joint distribution is evident. Joint distributions characterized by well-developed tails indicate a high probability of joint occurrence of extremely small and/or large values. In some practical applications, it is useful to have an estimate of this probability. Given the margins $F_1(x_1)$ and $F_2(x_2)$ and the copula $C(u_1,u_2)$,  the coefficients of lower and upper tail dependence provide such an estimate and are defined as
\begin{align}
\lambda_l & = \lim_{t \to 0^{+}} \Pm (x_2 \leq F_2^{-1}(t) | x_1 \leq F_1^{-1}(t)); \nonumber \\
& = \lim_{t \to 0^{+}} \frac{C(t,t)}{t},
\end{align}
respectively,
\begin{align}
\lambda_u & = \lim_{t \to 1^{-}} \Pm (x_2 > F_2^{-1}(t) | x_1 > F_1^{-1}(t)); \nonumber \\
& = \lim_{t \to 1^{-}} \frac{1 - 2t + C(t,t)}{1-t}.
\end{align}
These coefficients are conditional probabilities that measure the tendency of the random variable $x_2$ to behave as the random variable $x_1$. When their value is close to one, it means tail dependence (i.e., high probability of joint extreme values), when close to zero it means tail independence (i.e., low probability of joint extreme values). The analytical expression of $\lambda_l$ and $\lambda_u$ is available for various parametric copulas. For instance, the random points in the bottom-left panel of Fig.~\ref{fig:fig_ktau} has been generated through a Clayton copula with coefficient $\theta=5,$ for which $\lambda_l=0.87$ and $\lambda_u=0$. These values also hold for the other distributions in the same figure. 

\end{document}